\documentstyle[12pt]{article}

\def\ai{\'{\i}}
\def\itt{\int_{\tau_1} ^{\tau_2}}
\def\qb{\overline Q}
\def\pp{\pi_\phi}

\def\po{\pi_\Omega}
\def\om{\Omega}
\def\33{e^{3\Omega}}
\def\66{e^{6\Omega}}

\def\e-3{e^{-3\Omega}}

\def\be{\begin{equation}}
\def\ee{\end{equation}}

\def\pb{\overline P}
\begin{document}

\baselineskip.33in

\centerline{\large{\bf Comment on the transition amplitude for toy universe models}}

\bigskip

\centerline{Hern\'an De Cicco$^{a,}$\footnote{Electronic address: decicco@cnea.gov.ar} and Claudio Simeone$^{b,c,}$\footnote{Electronic address: simeone@tandar.cnea.gov.ar}}

\bigskip

\noindent {\it a) Centro At\'omico Constituyentes,  Comisi\'on Nacional de Energ\ai a At\'omica, Av. del Libertador 8250 - 1429 Buenos Aires, Argentina.}

\noindent {\it b) Departamento de F\ai sica, Comisi\'on Nacional de Energ\ai a At\'omica,
 Av. del Libertador 8250 - 1429 Buenos Aires, Argentina.}

\noindent {\it c) Departamento de F\ai sica, Facultad de Ciencias Exactas y Naturales, Universidad de Buenos Aires,  Ciudad Universitaria, Pabell\'on I - 1428, Buenos Aires, Argentina.}

\vskip1cm

ABSTRACT

\bigskip

The parametrized system called ``ideal clock'' is turned into an ordinary gauge system and quantized by means of a path integral in which canonical gauges are admissible. Then the  possibility of applying the results to obtain the transition amplitude for empty minisuperspaces, and the restrictions arising from the topology of the constraint surface,  are studied  by matching the models  with the ideal clock. A generalization to minisuperspaces with true degrees of freedom is also discussed.  

\vskip1cm


\newpage

I. INTRODUCTION

\bigskip

When the time is included among the canonical variables of a system and they are all given in terms of an arbitrary parameter, the resulting formalism is invariant under reparametrizations. We can have an ``already invariant'' system, like the gravitational field, or we can ``construct'' one by taking any system and parametrizing it. Here we exploit this feature to achieve a better understanding of the path integral quantization  for minisuperspace models.

Consider an isotropic and homogeneous Friedmann-Robertson-Walker (FRW) universe whose metric
is 
$$ ds^2=N^2d\tau^2-a^2(\tau)\left({dr^2\over 1-kr^2}+r^2d\theta^2+r^2sin^2\theta d\varphi^2\right).$$

The action for this cosmological model reads [1]
\be S=\itt\left( \pp\dot\phi+\po\dot\om-NH\right) d\tau\ee
where $\phi$ is the matter field, $\om\sim\ln a(\tau)$, $\pp$ and $\po$ are their conjugate momenta, and $N$ is a Lagrange multiplier enforcing the Hamiltonian constraint
\be H=G(\phi,\om)(\pp^2-\po^2)+v(\phi,\om ) \approx 0.\ee
The presence of a constraint reflects the reparametrization invariance of the formalism, namely, that the  evolution of the system is given in terms of the parameter $\tau$ which does not have physical significance [1,2,3,4].

In the case of an empty model $(\phi=0)$ we have the simple constraint
\be H=-G(\om)\po^2+v(\om ) \approx 0.\ee
This constraint can be obtained from that of the parametrized system called ``ideal clock'' by means of an appropriate canonical transformation [5]. The ideal clock is said to be a ``pure gauge'' system, as it has only one degree of freedom and one constraint; it is obtained by promoting the time $t$ to the status of a canonical variable and considering its evolution in terms of the parameter $\tau$. Its action is 
\be S=\itt\left(p_t \dot t-NH\right) d\tau ,\ee 
with the constraint
\be H=p_t -R(t) \approx 0. \ee
To obtain the constraint (3), $R(t)$ must be equal to $t^2$.

In order to obtain the transition probability for a minisuperspace described by (3) we shall turn the ideal clock into an ordinary gauge system [6,7,8] and calculate the transition amplitude for it with the usual path integral procedure of Fadeev and Popov. Then we shall study the quantum behaviour of the cosmological models by matching them with the ideal clock. The result will generalize what is obtained by setting $\phi=0$ in the models of ref. 8, and, mainly, will provide us with a  way to understand how the topology of the constraint surface restricts the gauge choice. We shall also briefly discuss the possibility of generalizing our procedure to models with true degrees of freedom.  

\vskip1cm

II. PATH INTEGRAL FOR THE IDEAL CLOCK

\bigskip

According to the usual path integral procedure to quantize gauge systems [3] (i.e. those with constraints that are linear and homogeneous in the momenta) in the case of only one constraint $G\approx 0$ the amplitude for the transition $|Q^i_1, \tau_1 >\ \to \ |Q^i_2,\tau_2>$ is given by

\be <Q^i_2,\tau_2 |Q^i_1,\tau_1 >=\int DQ^i DP_i DN \delta (\chi) |[\chi ,G]| e^{iS}\ee
where $\chi$ is a gauge condition which selects only one point from each class of physically equivalent points in phase space, and $|[\chi ,G]|$ is the Fadeev-Popov determinant, which makes the result independent of the gauge choice; it is clear that it must be $|[\chi ,G]|\neq 0.$ The fact that the constraint is linear and homogeneous in the momenta makes canonical gauge conditions  admissible in the path integral [6,7].
  
The ideal  clock described by the Hamiltonian (5) is a parametrized system, but we can turn it into an ordinary gauge system, in order to compute its quantum transition amplitude by means of the path integral (6) with $G=H$. To do this,  two succesive canonical transformations [7] are needed. The first transformation, $(t, p_t) \ \to \ (\qb^0,\pb_0),$ is generated by the function $W$ solution of the Hamilton-Jacobi equation 
\be {\partial W\over\partial t}-R(t)=E;\ee
matching $E=\pb_0$, a simple calculation gives 
\be W(t,\pb_0) =\pb_0\ t+\int R(t) dt, \ee
so that $\qb^0,$  $\pb_0$ and the new hamiltonian $\overline K$ are related to $t,$ $p_t$ and $H$ by
 \be \qb^0={\partial W\over\partial\pb}=t,\ \ \ \  p_t={\partial W\over\partial t}= \pb_0+R(t),\ \ \ \  \overline K=NH.\ee
The  variables $\qb^0$ and $\pb_0$  verify
$$[\qb^0,\pb_0]=[\qb^0,H]=1,$$
so that $\qb^0$ can be used to fix the gauge [3].

The second transformation is generated by
\be F=P_0\qb^0 +f(\tau),\ee
which yields
\be \pb_0={\partial F\over\partial \qb^0}=P_0,\ \ \ \ \  Q^0={\partial F\over\partial P_0}=\qb^0 , \ee
and a new non vanishing Hamiltonian
\be K=\overline K+{\partial f\over\partial\tau}=NP_0+{\partial f\over\partial\tau}\approx{\partial f\over\partial\tau}.\ee
Then, as a functional of $Q^0$ and $P_0$ the action of the ideal clock reads
\be S =\itt\left(  P_0{dQ^0\over  d\tau}-NP_0 -{\partial
f\over\partial\tau}\right) d\tau\ee
and in terms of the original variables [7]
\be S=\itt\left( p_t{dt\over d\tau}-N(p_t-R(t))\right)  d\tau + {[Q^0P_0-W-f]}_{\tau_1} ^{\tau_2} .\ee
The action (13) is that of an ordinary gauge system, as it has a non vanishing Hamiltonian ${\partial f\over\partial\tau}$ and a constraint $P_0\approx 0$ which is linear and homogeneous in the momenta. The additional terms in (14) do not modify the dynamics, as it is clear that they can be included in the integral as a total derivative. However, to guarantee that the new action weighs the paths in (6) in the same way that the original one does, we must choose $f$ so that these terms vanish in a gauge such that $\tau=\tau (t)$ [7]. With the canonical  gauge choice
\be \chi\equiv Q^0-\tau =t-\tau =0 \ee
we must choose 
\be f(\tau)=-\int R(\tau )d\tau .\ee 
From (15) we have $|[\chi, G]|=|[Q^0, H]|=1$, $\delta(\chi )=\delta(Q^0-\tau )=\delta(t-\tau ),$  so that the transition amplitude is
\begin{eqnarray}
 <t_2,\tau_2\vert t_1,\tau_1>&=&\int DQ^0 DP_0 DN\delta (Q^0- \tau) \exp\left( i{\itt\left[  P_0{dQ^0\over  d\tau}-NP_0 -{\partial
f\over\partial\tau}\right] d\tau } \right)\nonumber\\
&=&\int DQ^0 DP_0\delta(P_0)\delta (Q^0- \tau) \exp\left( i\itt\left[  P_0{dQ^0\over  d\tau} -{\partial
f\over\partial\tau}\right] d\tau \right)\nonumber\\
&=&  \exp\left(i\itt    R(\tau)    d\tau
\right). \end{eqnarray}
Hence, the probability for the transition from $t_1$ at $\tau_1$ to $t_2$ at $\tau_2$ is
\be \vert<t_2,\tau_2|t_1, \tau_1>\vert ^2\ =\ 1\ee
for any values of $t_1$ and $t_2$. This just reflects that the system has no true degrees  of freedom, because given $\tau$ we have only one possible $t$. We should emphasize that even though we have used a gauge which makes this fact explicit, the path integral is gauge invariant, and then we could have computed it in any gauge and the result  would have been the same. This can easily be verified by, for example, calculating the path integral in terms of the original variables with the action (14) and the canonical gauge choice $\chi\equiv t=0$:
\begin{eqnarray} 
<t_2,\tau_2\vert t_1,\tau_1>& =& \int Dt Dp_t  DN \delta (\chi ) |[\chi,H]|\times\nonumber\\
&  & \mbox{}\times \exp\left( {i \itt 
\left[ p_t  {dt  \over d\tau} -NH \right] d\tau-i\int_{t(\tau_1)} ^{t(\tau_2)}R(t) dt +i\itt
 R(\tau ) d\tau }\right)\nonumber\\ 
& = & \int Dt Dp_t   \delta (\chi )\delta (p_t-R(t)) |[\chi,H]| \times\nonumber\\
& &\ \ \ \ \ \mbox{}\times \exp\left({i \itt 
p_t  {dt  \over d\tau} d\tau-i\int_{t(\tau_1)} ^{t(\tau_2)}R(t) dt +i\itt
 R(\tau ) d\tau }\right) \nonumber\\
&=& \exp\left({i\itt    R(\tau)    d\tau
}\right).\end{eqnarray}
\vskip1cm

III. EMPTY MINISUPERSPACES

\bigskip

A way to get a better understanding of the quantization of certain minisuperspace models is to recall that their Hamiltonian constraints can be obtained starting from a mechanical system which has been parametrized by including the time $t$ as a canonical variable. It can be shown that the Hamiltonian constraint for an empty minisuperspace
$$H=-G(\om)\po^2+v(\om ) \approx 0$$
can be obtained from that of the ideal clock with $R(t)=t^2,$
$$H=p_t-t^2\approx 0,$$
by means of a canonical transformation. If we define [5]
\be V(\om )=sign(v)\left( {3\over 2}\int\sqrt{|v|\over G} d\om\right)^{2/3}\ee
the canonical transformation is given by
\be \po= -t{\partial V(\om)\over\partial\om},\ \ \ \ \ \ 
p_t= V(\om ).\ee
On the constraint surface $p_t-t^2=0$ we obtain
\be t= \pm V (\om), \ee 
and then, we can try to quantize the minisuperspace by means of a path integral in the variables $t,p_t$. This clearly depends  on the existence of a relation
$$V(\om)\leftrightarrow\om ,$$
but, as we shall see, the main restriction will be given by the geometrical properties of the constraint surface.

The most general form of the potential for an empty FRW minisuperspace is
\be v(\om) = -k e^\om + \Lambda\33 .
\ee
Let us first consider  the simple models with $k=0$ or $\Lambda=0$. For $k=0$ (flat universe, non zero cosmological constant) we have
$$v (\om)=\Lambda e^{3\om} ,$$
and for $\Lambda=0$, $k=-1$ (null cosmological constant,  open universe) we have the potential
$$v (\om)= e^{\om}.$$
In both cases, as well as for the open $(k=-1)$ model with non zero cosmological constant $\Lambda>0$
$$v(\om) =  e^\om + \Lambda\33 ,$$
 given $v$ and then $V(\om)$ we can obtain  $\om=\om(V)$ uniquely. As $\om\sim \ln a(\tau)$, from (22) we then see that in the simplest cases our procedure is basically equivalent to identifying the scale factor $a(\tau)$ of the metric with the time $t$  of the ideal clock. As in this cases the potential has a definite sign, the constraint surface splits into the two disjoint sheets
$$\po=\pm\sqrt{{v(\om)\over G(\om)}}.$$
Hence  the gauge fixation in terms of the coordinate $t$ of the ideal clock, which selects only one path in the $(t,p_t)$ phase space, also selects only one path in the $(\om,\po)$ phase space; this makes the quantization of this toy universes trivial, yielding a unity probability for the transition from $\om_1$ to $\om_2.$  

For the case $k=1$, $\Lambda > 0$ (closed model with non zero cosmological constant), the potential
$$ v(\om) = -e^\om + \Lambda\33 , $$
 is not a monotonic function of $\om$, but it changes its slope when
\be
\om=\ln \left({1\over \sqrt{3\Lambda}}\right)\ee
where it has a minimun, so that for a given value of $v(\om)$ we would have two possible values of $\om.$  However, physical states lie on the constraint surface 
$$ -G(\om)\po^2 -e^\om + \Lambda\33 = 0,$$   
which restricts the motion to

$$\po=\pm\sqrt{{e^\om(\Lambda e^{2\om}-1)\over G}}.$$ As $G$ is a positive definite function of $\om$, the condition that $\po$ must be real gives
\be
\om\geq\ln\left({1\over \sqrt\Lambda}\right).\ee
Hence,   the potential does not change its slope  on the constraint surface; this allows us to obtain  $\om=\om(v)$ and  the relation
$$\om={1\over 2}\ln\left({\Lambda^{2\over 3} V +1\over\Lambda}\right)$$
holds in  the physical phase space. There is, however, a problem resulting from the fact that the potential has not a definite sign: as $\po=0$ is possible in this model, the system can evolve from $\om_1$ to $\om_2$ by two paths. Then given a gauge condition in terms of $t$ we do not obtain a parametrization of the cosmological model in terms of $\om$ only, and then we cannot ensure that a  path integral for $t_1 \to t_2$ is equivalent to the path integral for $\om_1 \to \om_2$. This is  related to the fact that, differing from the previous examples, this model does not allow for the existence of an intrinsic time (see section V).

\vskip1cm

IV. TRUE DEGREES OF FREEDOM

\bigskip

The interest of systems with only one degree of freedom is clearly more conceptual than practical. One should try to consider more degrees of freedom starting from the action [10]
\be S[q^i,p_i,N]  = \itt\left[ p_t{dt\over d\tau}+p_k{dq^k\over d\tau}-N\left( p_t +H_0(q^k, p_k)-R(t)\right)\right] d\tau ,\ee
where
$H=p_t+H_0 -R(t) \approx 0$, which is obtained by including the time $t$ among the canonical coordinates of a mechanical system and recalling that a total derivative of $t$ does not change the dynamics. A very simple example in which this idea can be easily applied is that of the Hamiltonian constraint 
\be H={1\over 4}e^{-3\om}(\pp^2-\po^2)+e^\om\approx 0,\ee
which corresponds to a FRW minisuperspace model with a massless scalar field $\phi$ and non zero curvature $k=-1$. The potential has a definite sign, so that we shall not find the obstruction just discussed for the potential $-e^\om+\Lambda e^{3\om}$. It is easy to see that by applying our procedure we are able to quantize this system avoiding derivative gauges [6,7,1]:  the  parametrized system given by the action (26) is turned into an ordinary gauge system by a generalization of the canonical transformation  used for the ideal clock, and then canonical gauges are admissible [7]. In fact, we can impose a canonical gauge condition which identifies the coordinate $\om$ with  a monotonic function $T(\tau)$, so that $\om$ plays the role of a global  phase time (strictly speeking, as a global phase time $T$ must fulfill $[T,H]>0,$ we must choose $\om=T$ in the sheet $\po<0$ of the constraint surface, and  $\om=-T$ in the  sheet $\po>0$). Then, after integrating on the multiplier $N$ and on the pure gauge variables $Q^0$ and $P_0$ (see ref. 7)    we obtain the transition amplitude
\be <\phi_2,\Omega_2\vert \phi_1,\Omega_1>=\int  DQDP\exp\left(i\int_{T_1} ^{T_2}
PdQ\pm\sqrt{P^2+4 e^{4T}} dT\right) \ee
where the boundaries are given by $T_1=\om_1$, $T_2=\om_2$, and the paths  in phase space go from $Q_1=\phi_1$ to $Q_2=\phi_2$. The result shows the separation between physical degrees of freedom ($\phi$) and time ($\om$). The reduced system is governed by a time-dependent true Hamiltonian; this reflects that the field $\phi$ evolves subject to changing ``external'' conditions, the metric which plays the role of time. 

The  expression (28) makes simple to compute the infinitesimal propagator (to obtain the finite propagator we should still  integrate on $Q$).  As the path integral (28) is analogous to that for a ``relativistic particle'' with a ``mass'' $m=2e^{2T}$  in Minkowski space, by recalling the results of ref. 11 (see equation (68)) with $\nu=\mp 1$, $\gamma=1$ and $\sigma={1\over 2}\sqrt{(T_2-T_1)^2-(Q_2-Q_1)^2}$ we obtain
\be
 <\phi_2,\om_1+\epsilon \vert \phi_1, \om_1>=\pm{\epsilon\  e^{2\om_1}\over \sqrt{\epsilon^2-(\phi_2-\phi_1)^2}}H_1^{(1)}(2e^{2\om_1}\sqrt{\epsilon^2-(\phi_2-\phi_1)^2}),\ee
where $H_1^{(1)}$ is the Hankel function defined in terms of the Bessel functions $J_1$ and $N_1$ as $H_1^{(1)}=J_1+iN_1$. This propagator fulfills the boundary condition $<\phi_2,\om_1+\epsilon \vert \phi_1, \om_1>\ \to\ \delta(\phi_2-\phi_1)$ when $\epsilon \to 0.$

We should emphasize that we have succeeded in quantizing the model by imposing a canonical gauge condition because the potential has a definite sign and then allows to parametrize the system in terms of the coordinate $\om$; this is not the general case. 

\vskip1cm

V. DISCUSSION

\bigskip

The gauge choice is not only a way to avoid divergences in the path integral for a constrained system, but also a reduction procedure to physical degrees of freedom. When we choose a gauge to perform the integration in (6), at each $\tau$ we select one point from each class of equivalent points; if we do this with a system which is pure gauge, i.e. that has only one  degree of freedom and one constraint, we select only one point of the phase space at each  $\tau .$ For example, the gauge choice (15), $t-\tau=0,$
means that the paths in the phase space can only go from $t_1=\tau_1$ at $\tau_1$ to $t_2=\tau_2$ at $\tau_2$; there is no other possibility. Hence, the probability that the system evolves from $t_1$ at $\tau_1$ to $t_2$ at $\tau_2$ cannot be anything else but unity. Then if we can write the time $t$ in terms of only the coordinate $\om$ of a minisuperspace, its evolution is parametrized in terms of $\om$, there is only one possible value of $\om$ at each $\tau$, and the quantization of the model is therefore trivial.

It is clear that the models studied in the present work have an almost purely formal interest.
However, by matching them with the ideal clock they are   useful  to get a better  understanding of some of the difficulties that we find when we try to quantize the gravitational field.    
In particular, our procedure reflects the obstruction that a potential with a non definite sign can mean for the path integral quantization if one wants to impose canonical gauges: when the constraint hypersurface does not split into two disjoint sheets, we cannot find a function of the coordinates only which plays the role of time coordinate;  this of course generalizes to systems with true degrees of freedom.

 Let us go back to the general potential for an empty model $v(\om)=-ke^\om+\Lambda e^{3\om},$ with  $k=\pm 1.$ As the natural size of the configuration space for the case $k=1$ is given by (25), from (20) with $G={1\over 4}e^{-3\om}$ we obtain
\be
V(\om)={\Lambda e^{2\om}-k\over \Lambda^{2/3}}\ee
and from (22) we have 
\be
t=\pm\Lambda^{-1/3}\sqrt{\Lambda e^{2\om}-k},\ee
with $k=\pm 1$. From (21) we have
\begin{eqnarray}
t & =& -\po\left({\partial V\over\partial\om}\right)^{-1}\nonumber\\
&=&-{1\over 2}\Lambda^{-1/3}e^{-2\om}\po,\end{eqnarray}
so that if we write the time in terms of $\om$ only, in (31) we must choose $+$ for $\po<0$ and $-$ for $\po>0$ to ensure that $[t,H]>0$ [9]. Now an important difference between the cases $k=1$ and $k=-1$ arises: for $k=-1$ the potential is positive definite for finite values of $\om$, so that $\po$ cannot change its sign; then, the system evolves on only one of two disjoint sheets, and given  the initial value of $\po$  the evolution can be parametrized by a function of $\om$. 
Instead, for $k=1$ the momentum can change its sign, so that the initial conditions do not suffice to define a time in terms of $\om$ only, and we must necessarily define $t=t(\om,\po)$ (extrinsic time [12]). This means a general restriction to our path integral procedure involving canonical gauges:  the nonexistence of a time in terms of the coordinates only (intrinsic time [12]) is equivalent to the impossibility of quantizing the system by imposing  a  gauge condition $\chi\equiv f(q^i, \tau)=0$ which gives
$\tau=\tau(q^i)$; then we cannot guarantee that the path integral in terms of the gauge system into which we turned a cosmological model  is equivalent to a path integral in the original variables (see section II and ref. 7).

\vskip1cm

REFERENCES

\bigskip

\noindent 1.  J.    J.   Halliwell, in {\it Introductory Lectures  on  Quantum
Cosmology,}  Proceedings  of  the Jerusalem Winter School on Quantum Cosmology
and Baby Universes, edited by  T. Piran, World Scientific, Singapore (1990).

\noindent 2.  P.  A.  M.   Dirac,  {\it Lectures  on  Quantum  Mechanics,}  Belfer
Graduate School of Science, Yeshiva University, New York (1964).

\noindent 3. M.  Henneaux and C.  Teitelboim, {\it Quantization of Gauge Systems,}
Princeton University Press, New Jersey (1992).

\noindent 4. A. O. Barvinsky, Phys. Rep. {\bf 230}, 237 (1993).

\noindent 5. S.  C.   Beluardi and R.  Ferraro, Phys. Rev. D {\bf 52}, 1963 (1995).

\noindent 6. M.   Henneaux, C.  Teitelboim and J.  D.  Vergara, Nucl. Phys. B 
{\bf 387}, 391 (1992).

\noindent 7. R. Ferraro and  C. Simeone,  J. Math. Phys. {\bf 38}, 599 (1997).

\noindent 8. C. Simeone, J. Math. Phys. {\bf 39}, 3131 (1998).

\noindent 9. P. H\'aj\ai cek, Phys. Rev. D {\bf 34}, 1040 (1986).

\noindent 10. R. Ferraro and D. Sforza, gr-qc/9808064, to appear in Phys. Rev. D.

\noindent 11. R. Ferraro, Phys. Rev. D {\bf 45}, 1198 (1992).

\noindent 12. K. V. Kucha\v r, in {\it Proceedings of the 4th Canadian Conference on General Relativity and Relativistic Astrophysics}, edited by  G. Kunstatter, D. Vincent and J. Williams, World Scientific, Singapore  (1992).

\end{document}